\documentclass{appolb}

\usepackage{epsfig}

\begin{document}

\pagestyle{plain}
\newcount\eLiNe\eLiNe=\inputlineno\advance\eLiNe by -1

\title{Transport calculations of $\bar p$-nucleus interactions}

\author{
A.B. Larionov$^{a,b,c}$\thanks{e-mail: larionov@fias.uni-frankfurt.de},
I.N. Mishustin$^{a,c}$, I.A. Pshenichnov$^{a,d}$,\\ L.M. Satarov$^{a,c}$,
and W. Greiner$^a$
\address{$^a$Frankfurt Institute for Advanced Studies, J.W. Goethe-Universit\"at,
         D-60438 Frankfurt am Main, Germany\\
         $^b$Institut f\"ur Theoretische Physik, Universit\"at Giessen,
             D-35392 Giessen, Germany\\ 
         $^c$Russian Research Center Kurchatov Institute, 
         123182 Moscow, Russia\\
         $^d$Institute for Nuclear Research, Russian Academy of Science, 
         117312 Moscow, Russia}
}

\maketitle

\begin{abstract}
The Giessen Boltzmann-Uehling-Uhlenbeck transport model is extended
and applied to the $\bar p$-nucleus interactions in a wide beam momentum
range. The model calculations are compared with the experimental data on 
$\bar p$-absorption cross sections on nuclei with an emphasis on extraction
of the real part of an antiproton optical potential. The possibility of 
the cold compression of a nucleus by an antiproton in-flight is also 
considered.
\end{abstract}


\section{Introduction}

Antiproton interactions with nuclei are of big interest since they
deliver the information on in-medium $\bar p N$ interactions 
\cite{Friedman07} related to the antiproton optical potential.
The real and imaginary parts of $\bar p$ optical potential close to 
the nuclear centre are still poorly known \cite{Nakamura84,Friedman05}.
As shown in \cite{Buer02}-\cite{ML08}, a deep enough real part,
$\mbox{Re}(V_{\rm opt})=-(150-200)$ MeV, might cause sizeable 
compressional effects in $\bar p$-doped nuclei.

Here, we present some results of the microscopic transport simulations
within the Giessen Boltzmann-Uehling-Uhlenbeck (GiBUU) model
\cite{GiBUU}. Our goal is twofold: (i) to determine the 
$\bar p$ optical potential by comparison with the data
on $\bar p$ absorption cross sections on nuclei; 
(ii) to evaluate the probability of $\bar p$ annihilation
in a compressed nuclear zone for the beam momenta 
0.1-10 GeV/c.

The model is described in Sec. II. In Secs. III and IV,
the calculations of $\bar p$ absorption cross section
on nuclei and of the dynamical compression of the $^{16}$O  
nucleus by moving antiproton are presented. 
The results are summarized in Sec. V.

\section{The GiBUU model}

The GiBUU model \cite{GiBUU} solves the coupled set of kinetic
equations for different hadrons ($N,~\bar N,~\Delta,~\bar\Delta,~\pi ...$).
These equations describe the time evolution of a system 
governed by the two-body collisions and resonance decays.
The Pauli blocking for the nucleons in the scattering final
states is accounted for. Between the two-body collisions,
the particles propagate according to the Hamiltonian-like
equations in the mean field potentials defined by the (anti)%
baryon densities and currents.

We use the GiBUU model in the relativistic mean field mode.
The single-particle energies $\epsilon_i=V_i^0+\sqrt{{\bf p}_i^{\star2}+m_i^{\star2}}$ 
of the nucleons ($i=N$) and antinucleons ($i=\bar N$) 
depend on the effective mass $m_i^\star=m_i+S_i$, where
$S_i=g_{\sigma i}\sigma$ is the scalar field, kinetic
three-momentum ${\bf p}_i^\star={\bf p}-{\bf V}_i$, 
and vector field
$V_i^\mu = g_{\omega i} \omega^\mu + g_{\rho i} \tau^3 \rho^{3\mu}
+ e_i A^\mu$. Here, $\sigma,~\omega,~\rho$ and $A$ are, respectively,
the isoscalar-scalar, isoscalar-vector and isovector-vector meson fields,
and the electromagnetic field.

The meson-nucleon coupling constants $g_{\sigma N},~g_{\omega N}$ and $g_{\rho N}$
are taken from the NL3 set of parameters \cite{LKR97} of a non-linear Walecka model.
The meson-antinucleon coupling constants are motivated by the $G$-parity transformation,
however, allowing for their strengths to be rescaled by a factor $0 <  \xi  \leq 1$
as $g_{\sigma \bar N} = \xi g_{\sigma N},~g_{\omega \bar N} = -\xi g_{\omega N}$
and $g_{\rho \bar N} = \xi g_{\rho N}$\footnote{Pure $G$-parity transformed nuclear fields
are obtained with $\xi=1$. But this leads to an unrealistically deep real part of
the $\bar p$ optical potential, ${\rm Re}(V_{\rm opt})\simeq-660$ MeV.}.

The following two-body collision processes involving an antinucleon 
are implemented in the model: elastic scattering and charge exchange 
$\bar N N \to \bar N N$, inelastic production 
$\bar N N \to \bar B B + {\rm mesons}$, 
and annihilation $\bar N N \to {\rm mesons}$.
Further details of the model can be found in 
\cite{LMSG08,GiBUU,Blaettel93,Gaitanos,LPMG09} and in refs. therein.

\section{Antiproton absorption and annihilation on nuclei} 

\begin{figure}[t]
\unitlength1cm
\begin{picture}(6.,6.)
\put(1.0,0.0){\makebox{\epsfig{file=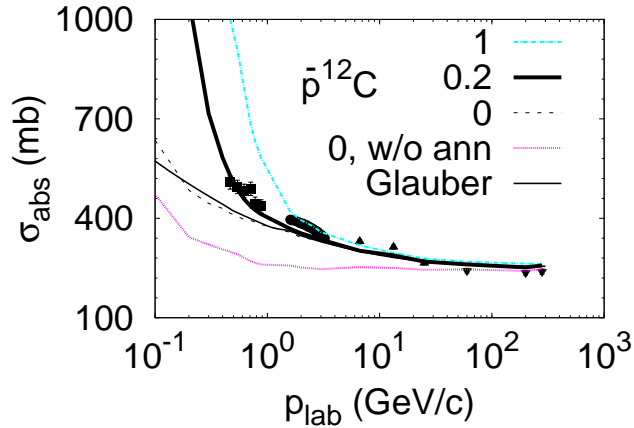,width=9.0cm}}}
\end{picture}
\caption{(Color online) Beam momentum dependence of the antiproton absorption cross section
on $^{12}$C. GiBUU calculations are represented by the curves denoted by the value of 
the scaling factor $\xi$ of meson-antibaryon coupling constants.
The results obtained using the Glauber model \cite{Glauber} are shown by thin
solid line. The calculation with $\xi=0$ without annihilation is shown by dotted line.
Experimental data are from \cite{Nakamura84}.}
\label{fig:sigabs_C12}
\end{figure}
Fig.~\ref{fig:sigabs_C12} shows the antiproton absorption
cross section on $^{12}$C.
The GiBUU calculation without nuclear part of the antibaryon mean field ($\xi=0$) is
in a very close agreement with the Glauber model prediction \cite{Glauber}\footnote{
Except for momenta less than $\sim0.2$ GeV/c, where the Coulomb potential causes 
the deviation.}.

The attractive real part of an antiproton optical potential
\begin{equation}
   \mbox{Re}(V_{\rm opt}) = S_{\bar p} + V_{\bar p}^0 
                          + \frac{S_{\bar p}^2-(V_{\bar p}^0)^2}{2m_N}   \label{Re_Vopt}
\end{equation}
bends the $\bar p$ trajectory towards the nuclear centre and increases the $\bar p$ absorption
cross section. One can see from Fig.~\ref{fig:sigabs_C12}, that the finite values of
the scaling factor $\xi\simeq0.2$ are needed to describe the antiproton absorption
cross section data at $p_{\rm lab}=470-880$ MeV/c measured at KEK \cite{Nakamura84}.
The best fit of the data \cite{Nakamura84} on $\bar p$ absorption cross sections on
the $^{12}$C, $^{27}$Al and $^{64}$Cu nuclei by GiBUU calculations is reached with
$\xi \simeq 0.21\pm0.03$, which produces $\mbox{Re}(V_{\rm opt}) \simeq -(150\pm30)$ MeV
in the nuclear centre. 

The imaginary part of the $\bar p$ optical potential can be calculated as
\begin{equation}
   \mbox{Im}(V_{\rm opt}) = -\frac{1}{2} 
                             <v_{\bar p N} \sigma_{\rm tot}^{\rm med}> \rho_N~, 
                                                    \label{Im_Vopt}
\end{equation}
where $v_{\bar p N}$ is the relative velocity of the antiproton and a nucleon;
$\sigma_{\rm tot}^{\rm med}$ is the total $\bar p N$ cross section including 
the in-medium effect of Pauli blocking for the final state nucleon in the
$\bar N N \to \bar N N$ channel; $\rho_N$ is the local nucleon density.
The averaging in Eq.(\ref{Im_Vopt}) is taken with respect to the nucleon
Fermi distribution. Using (\ref{Im_Vopt}), we obtain 
$\mbox{Im}(V_{\rm opt})\simeq-(105\pm5)$ MeV in the nuclear centre,
where a small uncertainty is due to different considered nuclei.
More details on the extraction of an
antiproton optical potential by GiBUU calculations can be found in 
\cite{LPMG09}.

\section{Dynamical compression of a nucleus by an antiproton}

\begin{figure}[t]
\unitlength1cm
\begin{picture}(10.,3.)
\put(0.0,0.0){\makebox{\epsfig{file=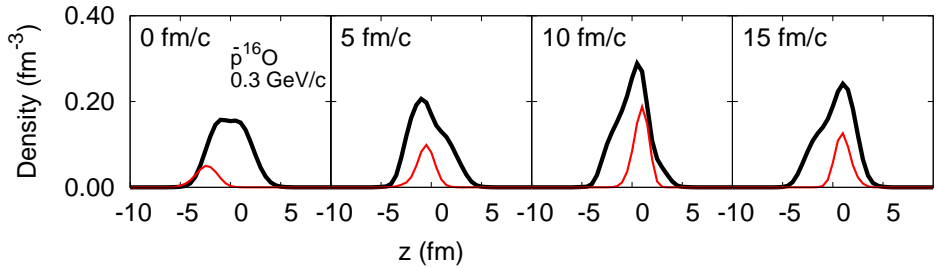,width=13.0cm}}}
\end{picture}
\caption{(Color online) Nucleon (thick solid lines) and antiproton
(thin solid lines) densities along the $z$-axis passing through
the nuclear centre $(x,y,z)=(0,0,0)$ for the $\bar p ^{16}$O system at
different times. The antiproton has been initialized at the 
coordinates $(0,0,-2.5)$ fm with momentum components $(0,0,0.3)$ GeV/c.
The scaling factor of the antiproton-meson coupling constants $\xi=0.22$ 
is used. Annihilation is switched-off.}
\label{fig:rhoz_ML}
\end{figure}
Fig.~\ref{fig:rhoz_ML} shows the time evolution of nucleon and antiproton
densities for the case of an antiproton initialized at the nuclear
periphery with momentum of 0.3 GeV/c directed towards the nuclear centre.
The calculation has been done without the $\bar N N$ annihilation channel
in the collision term, but allowing for the $\bar N N$ and $N N$ scattering.
We observe that the antiproton attracts surrounding nucleons catching them
into a potential well of about $70-100$ MeV depth. The nucleon density
bump moves with the $\bar p$ or slightly behind it due to some delay in the
reaction of the nucleon density on the perturbation created by the
antiproton.

The compression process of Fig.~\ref{fig:rhoz_ML} has to be complemented 
with the calculation of an antiproton survival probability 
\begin{equation}
   P_{\rm surv}(t) = \exp\left\{-\int\limits_0^t\,dt^\prime\,\Gamma_{\rm ann}(t^\prime)\right\}~.
                                               \label{Psurv}
\end{equation}
where $\Gamma_{\rm ann}= \rho_N \langle v_{\bar p N} \sigma_{\rm ann} \rangle$
is the antiproton annihilation width and $\sigma_{\rm ann}$ is the $\bar p N$
annihilation cross section (other notations are the same as in Eq.(\ref{Im_Vopt})).
In the case of the process shown in Fig.~\ref{fig:rhoz_ML},
the antiproton survives with the probability $P_{\rm surv}\sim10^{-3}$ at the time
10 fm/c when the nucleon density maximum is reached.
The experimental datection of the nuclear compression by the antiproton
is only possible, if $\bar p$ annihilates in the compressed zone of a nucleus.
The corresponding probability is
$P_{\rm compr}(\rho_c) = P_{\rm surv}(t_1) - P_{\rm surv}(t_2)$, 
where $t_1 < t_2$ are the times delimiting the time interval, when the
maximum density of a nuclear system exceeds some preselected value $\rho_c$.
In concrete calculations, we set $\rho_c=2\rho_0$ with 
$\rho_0=0.148$ fm$^{-3}$ being the nuclear saturation density. 

In the case of a real antiproton-nucleus collision, we apply, first,
the GiBUU model in the standard parallel ensemble mode to determine
the coordinates ${\bf r}$ and momentum ${\bf p}$ of the antiproton 
at its annihilation time moment event-by-event.
Next, we initialize $\bar p$ at $({\bf r},{\bf p})$ and run GiBUU 
without annihilation. This allows us to compute $P_{\rm compr}$ for a 
given annihilation event, \ie to determine the probability that this 
event will take place in a compressed nuclear zone. Finally,
we calculate the cross section of a $\bar p$ annihilation in
the compressed zone by weighting with the impact parameter as
$\sigma_{\rm compr}=\int_0^\infty db \, 2 \pi b \, 
\overline{P}_{\rm compr}(\rho_c,b)$, where the upper line denotes
averaging over annihilation events with a given impact parameter $b$.

\begin{figure}[t]
\unitlength1cm
\begin{picture}(6.,6.)
\put(2.0,0.0){\makebox{\epsfig{file=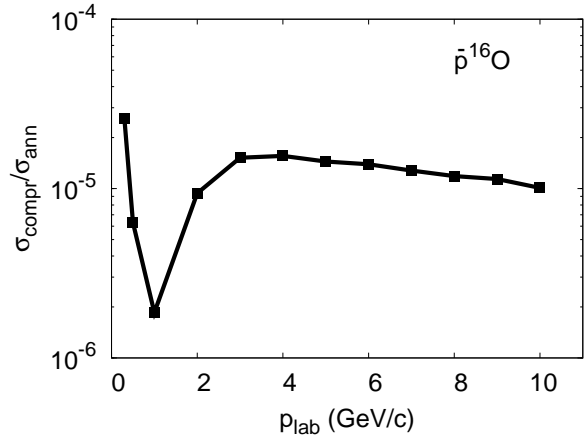,width=8.0cm}}}
\end{picture}
\caption{The probability of antiproton annihilation 
at the nucleon density exceeding $2\rho_0$ as a function of 
the beam momentum for $\bar p ^{16}$O collisions.} 
\label{fig:sigCompr_ML}
\end{figure}
In Fig.~\ref{fig:sigCompr_ML}, we present the probability
of $\bar p$ annihilation in a compressed zone given by the ratio 
$\sigma_{\rm compr}/\sigma_{\rm ann}$, where $\sigma_{\rm ann}$
is the total annihilation cross section of $\bar p$ on the
oxygen nucleus. The rise of the ratio with the beam momentum 
between about 1 and 3 GeV/c is caused by opening
the pion production channels $\bar N N \to \bar N N \pi (\pi...)$,
which leads to more intensive stopping of an antibaryon
before annihilation, and, therefore, to larger compression 
probabilities.

\section{ Summary and conclusions}
 
We have performed the microscopic transport GiBUU
simulations of $\bar p$-nucleus collisions. The antiproton
mean field potential has been described applying the non-linear
Walecka model supplemented by appropriate scaling 
of the meson-antinucleon coupling constants.

The extracted antiproton optical potential from comparison with 
the KEK data on $\bar p$ absorption
cross section below 1 GeV/c is $V_{\rm opt}=-(150\pm30)-i(105\pm5)$
MeV in the nuclear centre, which is comparable with the well known 
phenomenological values \cite{Nakamura84,Friedman05}.
However, the BNL and Serpukhov data on $\bar p$ absorption 
above 1 GeV/c require much deeper real part of about $-660$ MeV, 
close to the G-parity value. It is important, therefore,
to perform the new measurements of the $\bar p$ absorption cross
section on nuclei above 1 GeV/c at FAIR.

The probability of the antiproton annihilation in a compressed
nuclear zone is maximal at the beam momenta below 0.5 GeV/c.
However, it is also big enough ($\sim 10^{-5}$) at the beam momenta
from 3 to 10 GeV/c. The possibility of additional triggering, \eg on
a high-momentum proton, would prefer this beam momentum range with respect
to the low beam momenta for the study of annihilation events in the
compressed nuclear zone.

\begin{center}
                  Acknowledgments
\end{center}
This work was (financially) supported by the Helmholtz International
Center for FAIR within the framework of the LOEWE program (Landesoffensive 
zur Entwicklung Wissenschaftlich-\"Okonomischer Exzellenz) launched by the 
State of Hesse, by the DFG Grant 436 RUS 113/957/0-1 (Germany), and by the Grants 
NS-3004.2008.2 and RFBR-09-02-91331 (Russia).

\end{document}